\documentclass[twocolumn]{aastex631}
\usepackage[utf8]{inputenc}
\usepackage[autostyle]{csquotes}
\providecommand{\teff}{\ensuremath{T_{\rm eff}}}
\providecommand{\tprim}{T_{\rm 1}}
\providecommand{\tsec}{\ensuremath{T_{\rm 2}}}
\providecommand{\rprim}{\ensuremath{R_{\rm 1}}}
\providecommand{\rsec}{\ensuremath{R_{\rm 2}}}
\providecommand{\mprim}{\ensuremath{M_{\rm 1}}}
\providecommand{\msec}{\ensuremath{M_{\rm 2}}}

\providecommand{\msun}{\ensuremath{\,M_\Sun}}
\providecommand{\rsun}{\ensuremath{\,R_\Sun}}

\providecommand{\kms}{km s$^{-1}$}

\providecommand{\logg}{$\log{g}$}
\providecommand{\tic}{TIC-460388167}

\providecommand{\kms}{km~s$^{-1}$}

\usepackage{appendix}
\usepackage{placeins}
\usepackage{hyperref}
\usepackage{amsmath}
\usepackage{dirtytalk}
\usepackage{textcomp,gensymb}
\usepackage{footnote}
\usepackage{threeparttable}
\begin{document}

\title{Searching for GEMS: Discovery of the Nearby Post-Common-Envelope Binary System \tic\footnote{Based on observations obtained with the Hobby-Eberly
Telescope (HET), which is a joint project of the University of Texas at Austin, the Pennsylvania State University, Ludwig-Maximillians-Universitaet Muenchen, and Georg-August Universitaet Gottingen. The HET is named in honor of its principal benefactors, William P. Hobby and Robert E. Eberly.}}

\author[0009-0006-3467-630X]{Alexandra Boone}
\affiliation{Department of Physics \& Astronomy, University of Wyoming, Laramie, WY 82070, USA}

\author[0000-0002-4475-4176]{Henry A. Kobulnicky}
\affiliation{Department of Physics \& Astronomy, University of Wyoming, Laramie,
WY 82070, USA}

\author[0000-0003-4835-0619]{Caleb I. Ca\~nas}
\affiliation{NASA Goddard Space Flight Center, 8800 Greenbelt Road, Greenbelt, MD 20771, USA}

\author[0000-0001-8401-4300]{Shubham Kanodia}
\affiliation{Earth and Planets Laboratory, Carnegie Institution for Science,
5241 Broad Branch Road, NW, Washington, DC 20015, USA}

% \author[0000-0002-2990-7613]{Jessica Libby-Roberts}
% \affiliation{Department of Astronomy \& Astrophysics, The Pennsylvania State
% University, 525 Davey Laboratory, University Park, PA 16802, USA}
% \affiliation{Center for Exoplanets and Habitable Worlds, The Pennsylvania State University, 525 Davey Laboratory, University Park, PA 16802, USA}

\author[0000-0002-0048-2586]{Andrew Monson}
\author[0009-0003-7708-1347]{Peter Shea}
\affiliation{Steward Observatory, The University of Arizona, 933 N. Cherry Avenue, Tucson, AZ 85721, USA}

% \affiliation{Steward Observatory, The University of Arizona, 933 N. Cherry Ave, Tucson, AZ 85721, USA}

\author[0000-0001-9662-3496]{William Cochran}
\affiliation{McDonald Observatory and Department of Astronomy, The University of Texas at Austin}
\affiliation{Center for Planetary Systems Habitability, The University of Texas at Austin}

\author[0000-0001-9596-7983]{Suvrath Mahadevan}
\affiliation{Department of Astronomy \& Astrophysics, The Pennsylvania State University, 525 Davey Laboratory, University Park, PA 16802, USA}
\affiliation{Center for Exoplanets and Habitable Worlds, The Pennsylvania State University, 525 Davey Laboratory, University Park, PA 16802, USA}

% \author[0000-0003-4384-7220]{Chad Bender}
% \affiliation{Steward Observatory, The University of Arizona, 933 N. Cherry Avenue, Tucson, AZ 85721, USA}

% \author{Scott A. Diddams}
% \affiliation{Electrical, Computer \& Energy Engineering, University of Colorado, 1111 Engineering Dr., Boulder, CO 80309, USA}
% \affiliation{Department of Physics, University of Colorado, 2000 Colorado Avenue, Boulder, CO 80309, USA}

% \author[0000-0003-1312-9391]{Samuel Halverson}
% \affiliation{Jet Propulsion Laboratory, 4800 Oak Grove Drive, Pasadena, CA 91109, USA}

% \author[0000-0002-9082-6337]{Andrea S.J. Lin}
% \affiliation{Department of Astronomy, California Institute of Technology, 1200 E California Blvd, Pasadena, CA 91125, USA}

\author[0000-0001-8720-5612]{Joe Ninan}
\affiliation{Department of Astronomy and Astrophysics, Tata Institute of Fundamental Research, Homi Bhabha Road, Colaba, Mumbai 400005,
India}

% \author[0009-0006-7298-619X]{Varghese Reji}
% \affiliation{Department of Astronomy and Astrophysics, Tata Institute of Fundamental Research, Homi Bhabha Road, Colaba, Mumbai 400005,
% India}

\author[0000-0003-0149-9678]{Paul Robertson}
\author[0000-0002-7127-7643]{Te Han}
\affiliation{Department of Physics \& Astronomy, The University of California, Irvine, Irvine, CA 92697, USA}

%\author[0000-0001-9307-8170]{Brock A. Parker}
%\affiliation{Department of Physics \& Astronomy, University of Wyoming,
%Laramie, WY 82070, USA}

\author[0000-0001-8127-5775]{Arpita Roy}
\affiliation{Astrophysics \& Space Institute, Schmidt Sciences, New York, NY 10011, USA}
%\affiliation{Department of Physics and Astronomy, Johns Hopkins University,
%3400 N Charles St, Baltimore, MD 21218, USA}

\author[0000-0002-4046-987X]{Christian Schwab}
\affiliation{School of Mathematical and Physical Sciences, Macquarie University, Balaclava Road, North Ryde, NSW 2109, Australia}

\author{Madeleine Allen}
\affiliation{Department of Physics \& Astronomy, University of Wyoming, Laramie, WY 82070, USA}
% \author[0000-0001-7409-5688]{Gudmundur Stefansson}
% \affiliation{Anton Pannekoek Institute for Astronomy, University of Amsterdam, Science Park 904, 1098 XH Amsterdam, The Netherlands}

\begin{abstract}
Short-period white dwarf+main-sequence binaries are Post-Common-Envelope Binaries (PCEB) that have survived a common envelope phase. Such systems, if detached and eclipsing, enable precise measurements of the constituent stars, providing a unique opportunity to probe the effects of the common envelope phase on the system. We report the discovery of one such nearby (57 pc) system, TIC-460388167, using a combination of multi-band photometric light curves and spectroscopic radial velocities. In addition to eclipses, the system exhibits a continuously variable light curve that we model as a combination of ellipsoidal variations and star spots. We determine a period $P$=0.63596258$\pm$0.00000012 d and inclination $i$=89.0±0.4 deg. The best-fitting model specifies a white dwarf with $\tprim$=7607$\pm$127 K and radius $\rprim$=0.0131$\pm$0.0003 $R_\odot$, which is eclipsed by a $\tsec$=3151 $\pm$ 59 K, $\rsec$=0.327$\pm$0.006 $R_\odot$ M dwarf. The white dwarf mass is 0.61$\pm$0.04 M$_\odot$. We present the first velocity resolved profile for a PCEB secondary and show that the rotation of the M-dwarf is synchronous with the orbital period, as expected. We compare the constituent stars to other PCEB systems and find \tic A is one of the coolest known white dwarfs in such systems. \tic\ is among the longest period eclipsing PCEB systems known. 
\end{abstract}

\section{Introduction}
\label{sec:intro}
M dwarfs are the most common stars in the Milky Way. They are active targets of investigation to discover planets that are suitable for life. Their low luminosities allow for a habitable zone closer to the star relative to FGK dwarfs, making it more likely for transit or radial velocity methods to discover planets orbiting M dwarfs. Their smaller sizes result in deeper transit depths for a planet of fixed radius \citep{henry_character_2024, henry_solar_2006}. Additionally, M dwarfs have lower masses, which impart larger radial velocity signatures at a fixed planet mass. 

Targeted M-dwarf surveys, such as the M dwarfs Accompanied
by close-iN Giant Orbiters with SPECULOOS \citep[MANGOS,][]{Dransfield_2025arXiv251011528D}, Calar Alto high-Resolution search for M dwarfs with Exoearths with Near-infrared and optical Échelle Spectrographs \citep[CARMENES,][]{reiners_carmenes_2018}, Pervasive Overview of Kompanions of Every M-dwarf in Our Neighborhood \citep[POKEMON,][]{clark_pokemon_2022}, Search for Habitable Planets Eclipsing Ultra-cool Stars \citep[SPECULOOS,][]{delrez_speculoos_2018}, Next-Generation Transit Survey \citep[NGTS,][]{wheatley_next_2018}, and the \textit{Searching for Giant Exoplanets around M dwarf Stars} \citep[GEMS,][]{kanodia_searching_2024} are providing large sample of exoplanets to set tighter constraints on stellar and companion parameters and their occurrence rates. GEMS, in particular, is discovering large planets to reconcile observations and theories of planet formation. 
Current estimates of giant planet occurrence rates are uncertain \citep{Bryant_2023MNRAS.521.3663B, Gan_2023AJ....165...17G, glusman2025searchinggemsoccurrencegiant}, and the core-accretion theory of planet formation implies difficulty in forming giant exoplanets around smaller stars due to their less massive protoplanetary disks and longer Keplerian orbital timescales \citep{Laughlin_2004ApJ...612L..73L,andrews_mass_2013, pascucci_steeper_2016}. 
 
Surveys such as these often uncover astrophysical ``false positives'', e.g., eclipsing binaries, brown dwarfs, and triple systems \citep{Santerne_2012A&A...545A..76S,baroch_carmenes_2018,winters_spectroscopic_2020}, which can tend to be more common than actual planet detections \citep{glusman2025searchinggemsoccurrencegiant}. These systems are interesting in their own right, providing samples with high-quality datasets useful for determining low-mass star properties to better precision \citep{swayne_eblm_2024, kraus_mass-radius-rotation_2011} and constraining companion statistics. 
M dwarfs in particular need better measurements of fundamental stellar parameters, since they have shown large discrepancies from stellar models \citep{parsons_scatter_2018, Maxted_2023Univ....9..498M}. Due to the faint nature of M-dwarfs, many previous studies have been magnitude-limited, providing a biased sample compared with the recent volume-limited surveys. These more complete and unbiased datasets provide a parent sample that enables improved measurements of multiplicity \citep{clark_pokemon_2025, winters_solar_2019}, stellar activity \citep{huang_m-dwarf_2020, sethi_tight_2024}, and basic stellar parameters \citep{Maxted_2023Univ....9..498M}. 

% These surveys also open up the opportunity for a more detailed analysis on many systems.

One class of ``false positives'' being discovered with increasing frequency is binary systems consisting of a degenerate object (white dwarf; WD) and a main-sequence star (MS). Many of these are post-common-envelope binaries, which have survived a common-envelope phase, during which two stars orbited each other within a shared atmosphere resulting from the expansion of the more evolved star \citep{schreiber_age_2003,toonen_effect_2013}. The most common PCEB WD companion is a low-mass M-dwarf, with a mass distribution that peaks around 0.2 $M_\odot$ and falls steeply above 0.3 $M_\odot$ \citep{schreiber_post_2010, blomberg2024companionmassdistributionpost}. Although observational biases affect the characterization and  detection of such systems, Gaia-based studies such as \cite{santos-garcia_population_2025} also indicate that M-dwarfs are the most common companion. Although they are the most common type of PCEB, white dwarf/M-dwarf binaries are still rare, occurring at a rate of $\sim$0.04\% with respect to single stars \citep{smolcic_second_2004}. Short-period systems \citep[below three hrs,][]{rappaport_new_1983} are increasingly dominated by cataclysmic variables \citep[CVs,][]{warner_stellar_1976,warner_discovery_1995}, binary stars where the more evolved star accretes material from a Roche-lobe-filling main-sequence donor, although detached PCEBs also exist below this threshold \citep{nebot_gomez-moran_post_2011}. Above periods of three hours, such binaries are detached systems where the separation is too wide to be a CV. These detached systems are especially useful when eclipsing, as they enable the determination of stellar parameters with great precision \citep{law_three_2012, brown_characterising_2022}. Large catalogs containing over a thousand PCEB systems now exist, but still only a few dozen eclipsing systems are known \citep{rebassa-mansergas_magnitude-limited_2025, inight_towards_2021}.
Studies like these of both non-eclipsing and eclipsing PCEBs can contribute to the ongoing understanding of the common envelope phase \citep{zorotovic_post_2011, zorotovic_close_2022, torres_reconstructing_2025} and the impacts on its constituent stars \citep{toonen_effect_2013,camacho_monte_2014, cojocaru_population_2017, santos-garcia_population_2025}.

In this paper, we report the discovery of a detached eclipsing white-dwarf/M-dwarf system TIC-460388167 (components A and B, respectively), originally flagged as a possible planet in the GEMS survey for planets within 100 pc \citep{glusman2025searchinggemsoccurrencegiant} due to the deep and flat-bottomed eclipses in optical light curves.  Ground-based spectroscopic follow-up from the Habitable-zone Planet finder on Hobby-Eberly Telescope (HET) \citep{ramsey_1998SPIE.3352...34R, hill_2021AJ....162..298H} at McDonald Observatory \citep[HPF,][]{mahadevan_habitable-zone_2012, mahadevan_2014SPIE.9147E..1GM} showed large radial velocity signals that were inconsistent with a planetary system, leading to the system being distinguished as an eclipsing binary. We also utilize follow-up multi-band photometry performed at the Red Buttes Observatory \citep[RBO,][]{kasper_remote_2016}, ARCTIC at Apache Point Observatory \citep{evans_astrophysical_2016}, and the Three Hundred Millimeter Telescope \citep[TMMT,][]{Monson_2017AJ....153...96M} in order to more fully characterize this uncommon system. In Section \ref{sec:observations}, we present the data used in this analysis. In Section \ref{sec: analysis}, we focus on the methodology of determining the stellar and orbital parameters of the two stars. In Section \ref{sec:discussion}, we discuss the parameters of the system relative to larger sample of PCEBs. 

\begin{table}[h]
    \centering
    \caption{\tic\ Basic Parameters}
    \begin{tabular}{c c}
    \hline
    \hline
    Parameter & Value  \cr
    \hline   
    Gaia DR3 & 1156516327809897088 \cr
    i$^\prime$ (mag) & 15.5$\pm$0.01  \cr
    parallax (mas) & 17.627$\pm$0.033 \cr
    Distance (pc) & 57.1$\pm$0.3 \cr
    RA (2015.5) & 226.62878$\degree$ \cr
    Dec (2015.5)  & 4.66302$\degree$ \cr
    l ($\degree$) & 4.0629 \cr
    b ($\degree$) & +50.7549 \cr
    $\mu_{\rm RA}$ (mas \ yr$^{-1}$) &  -122.4$\pm$0.1 \cr
    $\mu_{\rm Dec}$ (mas \ yr$^{-1}$) & 21.7$\pm$0.2 \cr
    \hline
    \end{tabular}
    \label{tab: basic parameters}
\end{table}

\section{Observations} 
\label{sec:observations}
Table \ref{tab: basic parameters} lists the basic parameters for \tic\, including i$^\prime$ mag, parallax, distance, RA and Dec (2015.5), Galactic Long/Lat, proper motion from the NASA Exoplanet Archive.

Multi-color high-cadence photometry at several observatories revealed a chromatic flat-bottomed eclipse. Table \ref{Observation_Table} records the dates, cadence, pass band, and observatories used to obtain the photometric data.
%, establishing that TIC-460388167 contained two self-luminous objects instead of a planet.
\begin{table*}[!ht]
    \centering
    \caption{Summary of Photometry for TIC-460388167}
    \label{Observation_Table}
    \begin{longtable}{cccccc}
    \hline\hline
        BJD & BJD & Local Night & Exp. Time & Filter & Telescope \\ 
        (start) & (end) & (YYYY-MM-DD) & (s) &   &   \\ 
        \hline
        2459692.9671 & 2459717.5363 & 2022 & 600 & TESS T & TESS (S51) \\
        2460786.8314 & 2460786.9816 & 2025-04-18 & 45 & SDSS g$^\prime$ & APO (3.5 m) \\ 
        2460858.6525 & 2460858.8337 & 2025-07-01 & 45 & Bessel R & RBO (0.6 m) \\ 
        2460748.6639 & 2460748.8661 & 2025-03-13 & 120 & Bessel I & TMMT (0.3 m) \\ 
        2460762.6777 & 2460762.9012 &2025-03-27 & 120 & Bessel I & TMMT (0.3 m) \\
        2460785.5930 & 2460785.8040 &2025-04-19 & 120 & Bessel I & TMMT (0.3 m) \\ \hline
    \end{longtable}
        \addtocounter{table}{-1}
\end{table*}

\subsection{Multi-band Photometry}

\subsubsection{TESS Photometry}
\label{sec: TESS}

The Transiting Exoplanet Survey Satellite
\citep[TESS,][]{ricker_transiting_2014} observed TIC-460388167 with a 600~s cadence in Sector 51 from 2022 April 23 to 2022 May 18. We generated light curves using \texttt{Tess-Gaia Light Curves} \citep[tglc,][]{han_tessgaia_2023} package in \texttt{Python}, which models the effective point-spread function of TESS full-frame images \citep{FFI} and decontaminates the photometry using Gaia DR3 stellar information to avoid dilution of the light curve by nearby stars. There are, however, no other Gaia DR3 sources within 22 arcseconds of \tic.

\begin{figure*}
    \centering
    \includegraphics[width=1\linewidth]{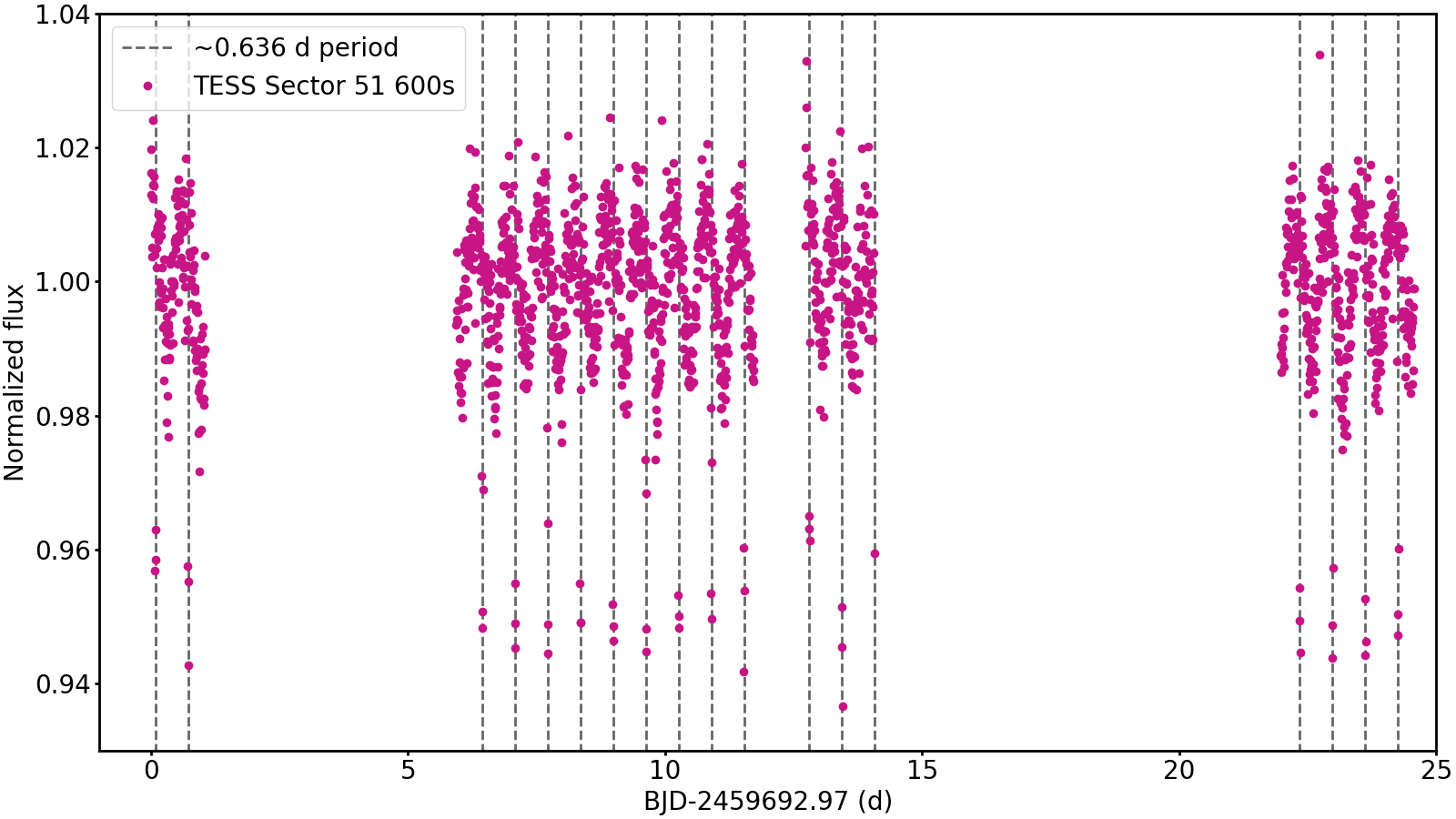}
    \caption{Full timeseries of TESS Sector 51 observations showing periodic dips every 0.636 days. Black dashed lines denote each visible dip. The light curve also exhibits additional out-of-eclipse modulations at the same period.}
    \label{fig:tessfullts}
\end{figure*}

The TESS dataset from Sector 51 spans $\sim$24 days. Figure \ref{fig:tessfullts} presents the full light curve, displaying periodic eclipses every $\sim$0.636 days, as marked by the vertical dashed lines. It also displays additional modulations with a 2\% semi-amplitude at the same period. Gaps in data are due to a combination of masked values from \texttt{tglc} data quality flags and download gaps from TESS.
% \FloatBarrier

\subsubsection{Apache Point Observatory SDSS g$^\prime$ Photometry}
\label{sec:APO}

\begin{table}[h]
    \centering
    \caption{\tic\ APO Photometric Data (This table is available in its entirety in machine-readable form in the online
article.)}
    \begin{tabular}{c c c}
    \hline
    \hline
    BJD$_{\rm{TDB}}$ & Normalized Flux & Flux Err  \cr
    \hline   
    2460786.83140 & 1.087 & 0.003 \cr
2460786.83196 & 1.09 & 0.003 \cr
2460786.83251 & 1.08 & 0.003 \cr
\nodata & \nodata & \nodata \cr
    \hline
    \end{tabular}
    \label{tab: APO table}
\end{table}

One eclipse of TIC-460388167 was observed in photometric conditions using the SDSS g$^\prime$ band on the local night of 2025 April 18 using ARCTIC \citep{evans_astrophysical_2016}, a general-purpose, visible-wavelength CCD camera on Apache Point Observatory's 3.5 m telescope. The ARCTIC plate scale is 0.45\arcsec\ pixel$^{-1}$ in 4$\times$4 binning mode.  The  instrument was slightly defocussed to an effective point spread function (PSF) of 3.2\arcsec\ full width at half maximum (FWHM).  ARCTIC captured the entire duration of one eclipse, which spanned 3.6 hours using 45-second exposures with a readout time of approximately 1.3 seconds, including pre- and post-eclipse baseline over an airmass range of 1.1--1.7. We calibrated the images using flat-fielding and bias subtraction, but dark current subtraction was not necessary for this exposure time. We extracted the corresponding light curve using AstroImageJ \citep[AIJ,][]{collins_astroimagej_2017} with a 4.5\arcsec\ aperture and a 7.65/11.25\arcsec\ inner/outer background annulus. The photometric RMS uncertainties are 0.2\%. We used 13 comparison stars and adopted the mid-exposure BJD$_\text{TDB}$ as computed by AstroImageJ, following \cite{eastman_achieving_2010}. Table \ref{tab: APO table} lists the mid-point times of each exposure, normalized flux, and error.
% 10-pixel aperture and a 17/25-pixel inner/outer background annulus.

\subsubsection{Red Buttes Observatory Photometry}
\label{sec: RBO}
\begin{table}[h]
    \centering
    \caption{\tic\ RBO Photometric Data (This table is available in its entirety in machine-readable form in the online
article.)}
    \begin{tabular}{c c c}
    \hline
    \hline
    BJD$_{\rm{TDB}}$ & Normalized Flux & Flux Err  \cr
    \hline   
    2460858.65982 & 0.991 & 0.026 \cr
2460858.66074 & 1.037 & 0.025 \cr
2460858.74402 & 0.977 & 0.017 \cr
\nodata & \nodata & \nodata \cr
    \hline
    \end{tabular}
    \label{tab: RBO table}
\end{table}
We observed TIC-460388167 in intermittently cloudy conditions with the 0.6 m University of Wyoming telescope at the Red Buttes Observatory \citep[RBO,][]{kasper_remote_2016} on the local night of 2025 July 1. We used the Bessel R band with 45-second exposures, $\sim$2-second readout time, and an additional $\sim$30 s deadtime per exposure for telescope repositioning. The telescope was equipped with an Apogee Alta F16 camera with Kodak KAF-16803 sensor, producing a field-of-view of 25.2\arcmin\ with a plate scale of 0.72\arcsec\ pixel$^{-1}$ in 2$\times$2 binning mode. The full duration of one eclipse was captured during the visit. We calibrated the raw data using dark, bias, and flat field corrections. We adopted BJD$_\text{TDB}$ mid-exposure times calculated from AIJ, and extracted the corresponding light curve using differential aperture photometry and 20 comparison stars with a 4.32\arcsec\ aperture and a 14.4/28.8\arcsec\ inner/outer background annulus. The typical seeing was $\sim$3.2\arcsec. The photometric RMS is 0.5\%. Table \ref{tab: RBO table} lists the mid-point exposure time, normalized flux, and uncertainty.
% 6-pixel aperture and a 20/40-pixel inner/outer background annulus.
\subsubsection{The Three Hundred Millimeter Telescope}
\label{sec: TMMT}
\begin{table}[h]
    \centering
    \caption{\tic\ TMMT Photometric Data (This table is available in its entirety in machine-readable form in the online
article.)}
    \begin{tabular}{c c c}
    \hline
    \hline
    BJD$_{\rm{TDB}}$ & Normalized Flux & Flux Err  \cr
    \hline   
    2460748.66393 & 1.018 & 0.033 \cr
2460748.66547 & 0.998 & 0.031 \cr
2460748.66701 & 1.044 & 0.031 \cr
\nodata & \nodata & \nodata \cr
    \hline
    \end{tabular}
    \label{tab: TMMT table}
\end{table}
TIC-460388167 was observed with The Three Hundred Millimeter Telescope \citep[TMMT,][]{Monson_2017AJ....153...96M} on three local nights from 2025 March to 2025 April in the Bessel I band with 120 s exposures. TMMT is a 0.3 m telescope with a 2048$^{2}$ pixel CCD detector that produces a 40.75\arcmin\ FOV with a pixel scale of 0.97\arcsec\ pixel$^{-1}$. We extracted one full eclipse on each night of observations, displayed in Table \ref{tab: TMMT table}. We calibrated the raw data using dark, bias, and flat field corrections, and adopted BJD$_\text{TDB}$ mid-exposure times calculated from AIJ. We extracted the corresponding light curve using differential aperture photometry in AIJ with a 3.88-pixel aperture and a 19.4/38.8-\arcsec\ inner/outer background annulus. The typical seeing was $\sim$4.3\arcsec\ on the first night in March and $\sim$2.9\arcsec\ on the other two nights. The photometric RMS was $\sim$1.5\% across all three nights. Table \ref{tab: TMMT table} provides the midpoint exposure times, normalized flux, and errors.
%4-pixel aperture and a 20/40-pixel inner/outer background annulus
\subsection{Spectroscopy}
% Radial velocities (RVs) were obtained using spectroscopic measurements of the system to determine the mass of the white dwarf and to characterize the activity level of the system. Table \ref{Tab:RV} lists the radial velocities and uncertainties. 

\subsubsection{LAMOST Spectrum}
\begin{figure}
    \centering
    \includegraphics[width=1\linewidth]{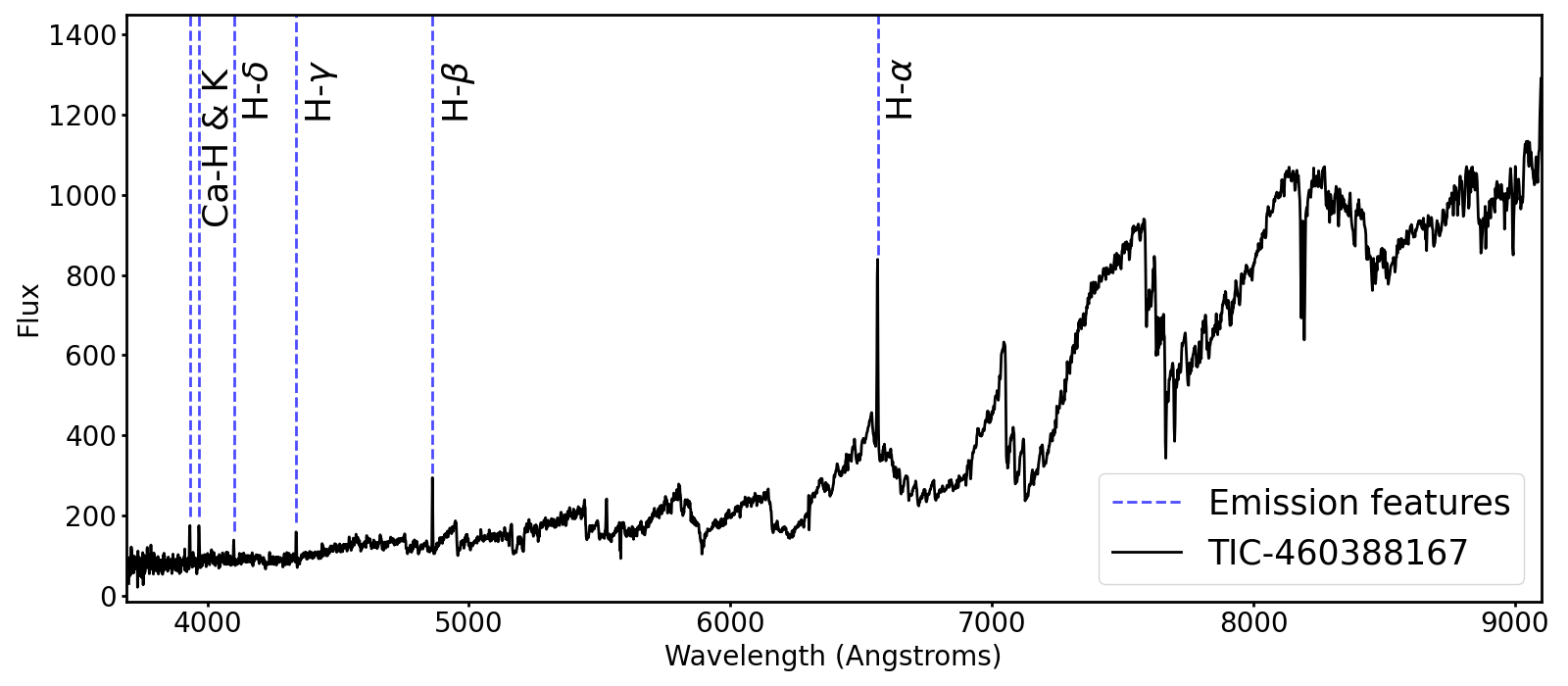}
    \caption{LAMOST Spectrum of \tic\ (\textit{black}) showing prominent emission features at Ca H\&K and in the Balmer series (\textit{vertical blue dashed lines}), indicating chromspheric activity.}
    \label{fig:lamostspectrum}
\end{figure}

We retrieved from the Large Sky Area Multi-Object Fiber Spectroscopic Telescope \citep[LAMOST,][]{zhao_lamost_2012} (also called the Guo Shou Jing Telescope; GSJT) archive (DR10 v2.0) a spectrum obtained on 2014 March 24 (JD=2456741.29727) with an exposure time of 4500 s as an aid in characterizing the activity of the system . The SNR was 58:1 per pixel around 8700 \AA. LAMOST covers 3700--9000 \AA\ wavelength range and has resolution R $\sim$1800 at 5500 \AA\ \citep{zhao_lamost_2012}. Figure \ref{fig:lamostspectrum} shows the LAMOST spectrum. Vertical dashed line mark the locations of Ca H\&K+ which appear in emission indicating chromospheric activity. The Balmer series also appears in emission. This spectrum is typical of a cool late-type star.

%\FloatBarrier

\subsubsection{HPF Spectra}
\label{sec:hpfspec}
\begin{table}[!ht]
\caption{Spectroscopic measurements from HPF}

\label{Tab:RV}
    \centering
    \begin{tabular}{ccc}
    \hline \hline
        $\text{BJD}_{\text{TDB}}$ & Bary. RV & $\sigma_{RV}$  
        \\
        \hspace{0.1cm} &  (km s$^{-1}$) &  (km s$^{-1}$) \\ 
         \hline
         2460739.89134 & 48.8   & 0.2 \\
         2460739.90280  & 64.8   & 0.2 \\
         2460734.90739 & 142.3  & 0.2 \\
         2460734.91866 & 144.2  & 0.2 \\
         2460762.96577 & 114.2  & 0.3 \\
         2460762.97761 & 104.1  & 0.2 \\
         2460821.67021 & -140.8 & 0.2 \\
         2460821.68173 & -149.1 & 0.2 \\
         2460735.90205 & -163.2 & 0.2 \\
         2460735.91339 & -157.2 & 0.2 \\
         2460744.87685 & -100.2 & 0.2 \\
         2460744.88821 & -84.7  & 0.2 \\
        \hline
    \end{tabular}
\end{table}
        % 2460739.88309 & 48.8   & 0.4 \\
        % 2460739.89505 & 64.8   & 0.3 \\
        % 2460734.89995 & 142.3  & 0.3 \\
        % 2460734.91192 & 144.2  & 0.3 \\
        % 2460762.95567 & 114.2  & 0.5 \\
        % 2460762.96764 & 104.1  & 0.4 \\
        % 2460821.66098 & -140.8 & 0.4 \\
        % 2460821.67295 & -149.1 & 0.4 \\
        % 2460735.89460 & -163.2 & 0.4 \\
        % 2460735.90657 & -157.2 & 0.4 \\
        % 2460744.86823 & -100.2 & 0.4 \\
        % 2460744.88019 & -84.7  & 0.4 \\  \hline
We obtained high-resolution spectra on six nights between 2025 February and 2025 May using the Habitable Zone Planet Finder (HPF, \cite{mahadevan_habitable-zone_2012}), a fiber-fed \citep{Kanodia_2018, Kanodia_2021ApJ...912...15K}, stabilized \citep{Stefansson_2016ApJ...833..175S} infrared echelle spectrograph on the 10 m Hobby-Eberly Telescope (HET) at the McDonald Observatory to determine radial velocity measurements of the system. HPF has wavelength coverage 8080--12780~\AA\ across 28 spectral orders and resolution R$\sim$55,000 \AA\ \citep[$\approx$6~\kms$\equiv$0.2 \AA\ at 10000,][]{hill_2021AJ....162..298H, ramsey_1998SPIE.3352...34R}. Six phase-spaced nights of two back-to-back 969 s exposures yielded 12 radial velocity (RV) data points, providing a full-phase RV curve. 

The data were processed using HxRGproc\footnote{\url{https://github.com/indiajoe/HxRGproc}} algorithms \citep{Ninan_2018SPIE10709E..2UN} and wavelength calibrated by the method described in \citet{stefansson_sub-neptune-sized_2020}. Routine laser frequency comb calibration frames allowed wavelength solutions to be determined with a drift correction to a precision of $<$30 cm s$^{-1}$ \citep{stefansson_sub-neptune-sized_2020}.  We removed spectral ranges strongly contaminated by telluric absorption. The typical signal-to-noise ratio of the extracted 1D spectra was 60:1 per 0.05 \AA\ pixel near 10,000 \AA.

The 28 spectral orders were continuum normalized and combined into a single spectrum. We converted the spectrum to a barycentric reference frame using \texttt{barycorrpy}\footnote{\url{https://github.com/shbhuk/barycorrpy}} \citep{kanodia_python_2018}, which is based on the code by \cite{wright_barycentric_2014}. We employed a custom python code to compute the broadening function \citep[i.e., similar to a cross-correlation function][]{rucinski_spectral-line_1992}, adopting a PHOENIX model atmosphere narrow-lined template \citep{husser_new_2013} corresponding to the temperature and $\log g$ of \tic B (see Section \ref{sec:specclass}). From a Gaussian fit to the broadening function peak we measured the stellar radial velocity. The typical radial velocity uncertainty on the Gaussian center is $\sim$0.2 \kms. Table \ref{Tab:RV} lists the Barycentric Julian dates at mid-exposure, RVs, and their associated uncertainties for each HPF exposure.

None of the spectra display Hydrogen Paschen emission. The expected stellar activity variability could lead to emission lines being present in one epoch that is not present in the HPF spectra. 
\section{Analysis}
\label{sec: analysis}
\subsection{Spectral Classification of the M dwarf}
\label{sec:specclass}
\begin{figure}
    \centering
    \includegraphics[width=1\linewidth]{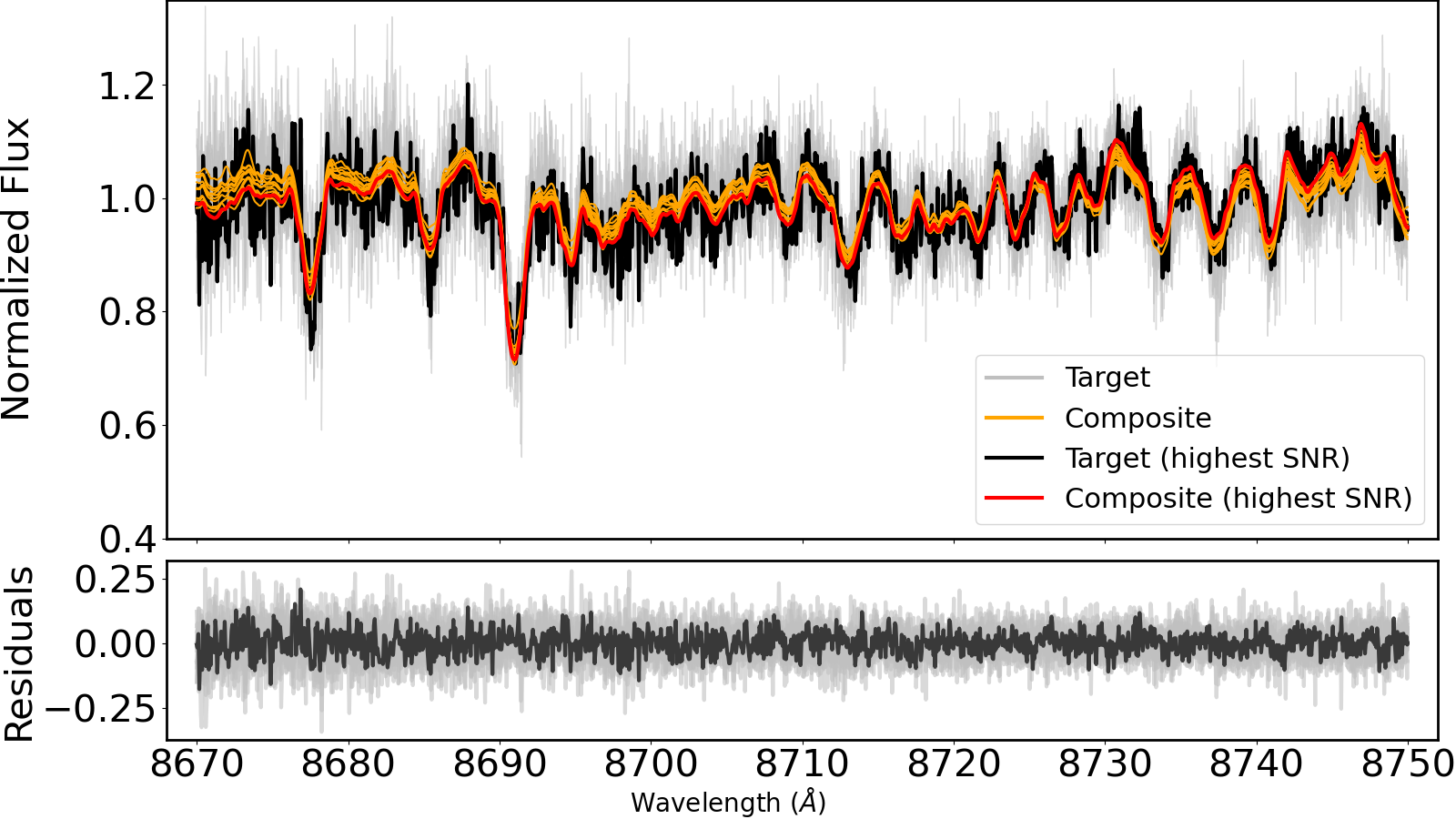}
    \caption{The \texttt{HPF-SpecMatch} spectra for order 5 {\textit{(top panel)}. The gray shows all the target spectra with each best-fitting composite in orange, and the black shows the highest S/N visit spectrum with its composite corresponding to \tsec=3151, [Fe/H] $=$ -0.044, and $\log g=5.0$ in red.  Residuals \textit{(bottom panel)}.}}
    \label{fig:hpfspecmatch}
\end{figure}
\FloatBarrier

We used the \texttt{HPF-SpecMatch}\footnote{\url{https://github.com/gummiks/hpfspecmatch}} package \citep{stefansson_neptune-mass_2023} to get initial stellar parameters from the spectrum obtained on 2025 May 26. \texttt{HPF-SpecMatch} uses a $\chi^2$ algorithm to search a  library of stellar templates and generate the best-fitting template.   It utilizes a catalog of stars that span 2700--4500 K, 4.6--5.3 $\log g$, and $-0.5$--$+0.5$ [Fe/H]. The five best-matching stars are combined into a composite spectrum to best characterize the target spectrum. 
We used two different spectral orders (order 5 and 17) because they had the least amount of telluric contamination in HPF's wavelength coverage. The derived values from order 5 show temperature $\tsec$$=$3151$\pm$59 K, $\log g=$ 5.03$\pm$0.04, and [Fe/H] $=$-0.044$\pm$0.016. Order 17 shows good agreement, with $\tsec$$=$3176$\pm$ 59 K, $\log g=$ 5.04$\pm$ 0.04, and [Fe/H] $=$-0.17$\pm$ 0.16. \footnote{The top six template matching stars for the order 5 and 17 spectra are GJ 3323, Ross 128, GJ 1289, GJ 1151, GJ 3258, and GJ 15 B.} As with previous results from the \textit{Searching for GEMS} survey \citep{kanodia_searching_2024,Reji_2025AJ....169..187R,Bernabo_2024AJ....168..273B}, we note that the complexities of M-dwarf spectra and limitations with the \texttt{HPF-SpecMatch} algorithm and range of the stellar library impede the accuracy of the derived metallicity.

Figure \ref{fig:hpfspecmatch} shows all the spectra in order index 5 for \tic\ in grey, and their composites from the best-fitting templates in orange. The spectrum corresponding to the hightest S/N is shown in black, with its corresponding composite in red. 
%The black line shows the spectrum with the highest S/N, and the composite of the five best-matching templates in red. 
%The highest S/N composite corresponded to a \tsec=3151, [Fe/H]=-0.044, and $\log g=5.03$. We used these values as priors in the later fitting of the M-dwarf.

\subsection{Light Curve}
\label{sec: lc}
We extracted the light curves of the three bandpasses using differential aperture photometry in AIJ and normalized the light curve by defining a baseline region outside of the eclipse. We fit a model \citep[\texttt{PHOEBEv2.4},][]{prsa_physics_2016} light curve to the four-passband light curve using a Markov Chain Monte Carlo (MCMC) sampling method with the \texttt{emcee} package \cite{foreman-mackey_emcee_2013}. We fixed the temperature of the M dwarf star at $\tsec \approx 3151\pm59 $ K based on the spectral analysis of the HPF spectrum. For the M-dwarf, we adopted the default limb and gravity darkening in PHOEBE, which are interpolated from PHOENIX atmosphere tables \citep{husser_new_2013}. For the WD, we used quadratic limb darkening coefficients from \cite{gianninas_limb-darkening_2013} for SDSS g$^\prime$, I, and R bands and \cite{claret_gravity_2020} for the TESS bandpass for the WD, assuming a hydrogen-dominated atmosphere. Free parameters were the temperature of the WD ($T_{\rm 1}$, modeled as a blackbody), the radius of the WD ($\rprim$), the radius of the M star ($\rsec$), inclination ($i$), period ($P$), and the time of mid-eclipse ($t_0$). We adopted loose Gaussian priors with $\sigma$ = 0.01 d on period and t$_0$, centered around the values found using a box least squares (BLS) periodogram on the TESS data. $\rprim$, $\rsec$, $\tprim$, inclination took on uniform priors centered around values found from an initial manual fit. The mass of $M_{\rm 2}$ was fixed at 0.34 \msun\ obtained from the \texttt{EXOFASTv2} fit to the spectral energy distribution to main-sequence evolutionary model (Section \ref{sec: rv}).  $M_{\rm 1}$=0.61 \msun\ and eccentricity ($e$=0.001) were fixed at values found from fitting the radial velocity data (Section \ref{sec: rv}).  We adopted an iterative approach to fit both the light curve data and radial velocity data self consistently. We accounted for the finite time of integration in all the data by using an oversampling method in PHOEBE to ensure the model correctly accounted for the exposure times and did not affect the extracted parameters. We considered effects of irradiation and reflected light but found that these effects were negligible ($\sim$0.015\%) relative to the typical photometric errors of the dataset. We do not include the effects of tidal distortion in the eclipse modeling, however, these effects are included in Section \ref{sec:outofeclipsevar} where we model the out-of-eclipse variations.
\begin{figure*}
    \centering
        \includegraphics[width=1.0\linewidth]{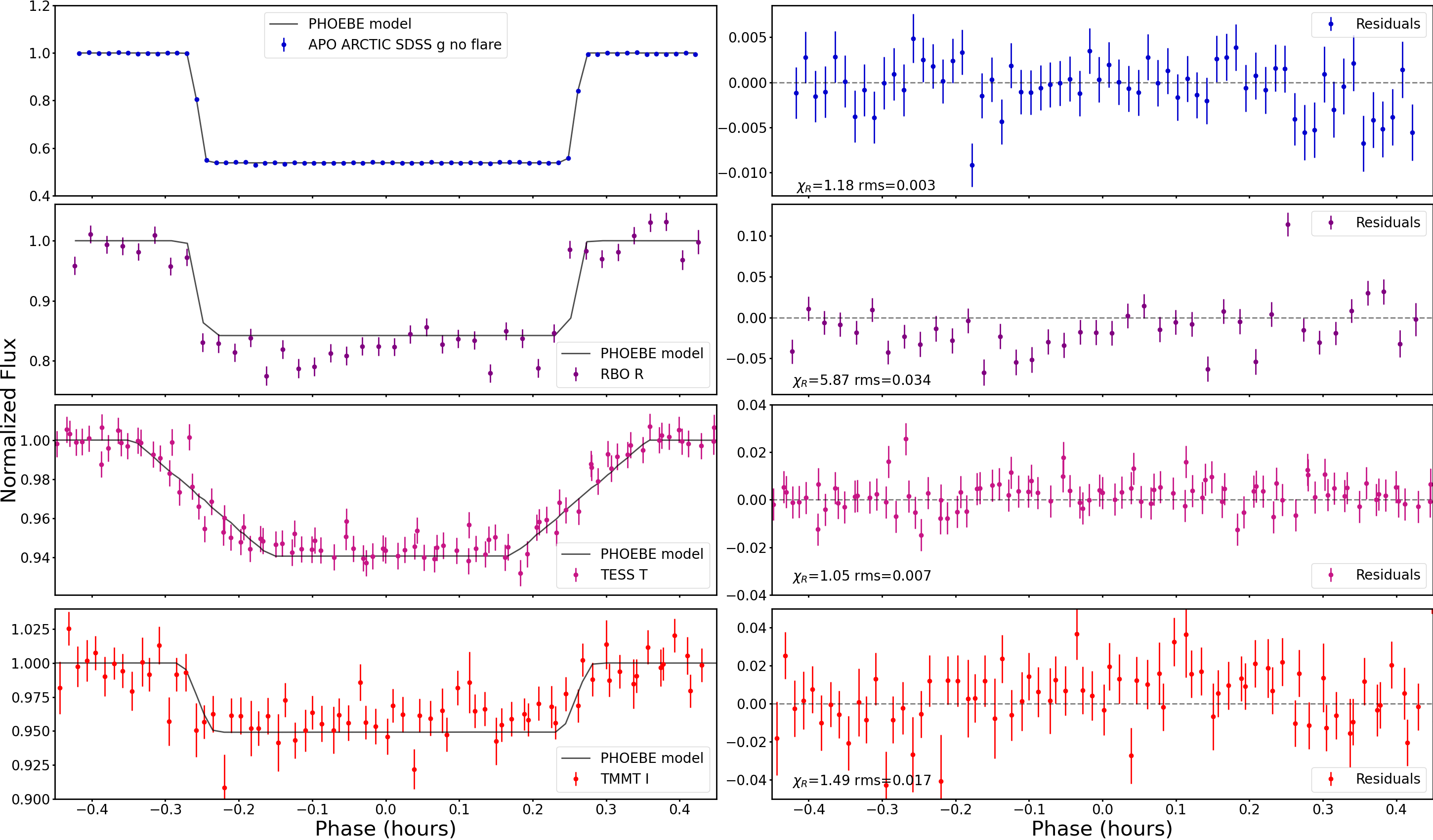}
    \caption{Light curves and best-fitting models (\textit{left column}) and residuals (\textit{right column}) for  SDSS g$^\prime$, Bessel R, TESS, and Bessel I bands (\textit{from top to bottom}). The x-axis displays orbital phase in hours centered on mid-eclipse. The strongly chromatic nature of the eclipse is evident.}
    \label{fig:4bandlc}
\end{figure*}
% \FloatBarrier

Figure \ref{fig:4bandlc} shows the g$^\prime$/R/TESS/I-band light curves of TIC-460388167. The data are normalized and phase-folded according to the best fitting period and $t_0$. The left column shows the folded data points, Poisson-based photometric uncertainties\footnote{The quoted uncertainties include Poisson noise from the source, background subtraction, and read noise.  {We adopt 0.01 as an average uncertainty on the TESS data, somewhat smaller than the uncertainties reported by {\tt tglc} which use an empirical RMS scatter in the data rather than Poisson-based statistics. This reduction was necessary to avoid $\chi^2_{\rm red}$$<$1 for the TESS light curve model fit. }}, and best-fitting \texttt{PHOEBE} model (\textit{black curve}). The right panels show the corresponding residuals for each dataset, with the reduced chi-squared and rms indicated on each panel. The y-axis scale is not uniform; the eclipse depths are chromatic, showing a nearly 50\% eclipse in the g$^\prime$ band that is only $\sim$5\% in Bessel I.  These multi-band data provide strong constraints on all of the modeled parameters, especially given the exquisite precision and time resolution of the g$^\prime$ measurements.   

Table \ref{tab: final parameters} reports the retrieved stellar parameters determined from the best-fitting model, including the temperature of the white dwarf and the radii of both stars. Table \ref{tab: orbital parameters} shows the retrieved orbital parameters, period, and $t_0$ from the light curve fitting. These represent the median value and a one-$\sigma$ confidence interval determined by the 16th and 84th percentile values of the MCMC posteriors. The posteriors for these parameters showed tightly constrained Gaussian distributions. The most probable period is $P = 0.63596258 \pm 0.00000012$ d, where the long ($>$2.99 yr) time baseline of the TESS and APO datasets provides very tight constraints on the period to a fraction of a second. The $t_0$ of 2460786.91153$\pm$0.00034 is chosen to lie at the time of the APO g$^\prime$ observations in 2025 April on account of the high quality of that dataset. The median inclination is 89.0$\pm$0.4 deg. The median $T_{\rm 1}$ is 7607$\pm$128~K, typical of a $\sim$1.2 Gyr white dwarf \citep{bedard_spectral_2020, Montreal_2017ASPC..509....3D}. The most probable radius of the white dwarf is $R_{\rm 1}$=0.0131$\pm$0.0004 $R_\odot$ or $\sim$1.43 $R_\oplus$, consistent with common white dwarfs in the mass range 0.5--0.7 $M_\odot$ \citep{pereiras_mass-radius_2025}. The radius of the M dwarf is \rsec\ = 0.327 $\pm$0.006 $R_\odot$, consistent with a mid-M dwarf \citep{pecaut_intrinsic_2013}. This radius is larger than the 0.22$\pm$0.03 \rsun\ radius predicted by the 3151 K temperature expected from standard temperature-radius relations \citep{Mann2015ApJ...804...64M, pecaut_intrinsic_2013}. The resolution to this discrepancy is not obvious. Most studies agree that the common-envelope stage does not affect the secondary star's mass or radius \citep{Hjellming_1991ApJ...370..709H}.  Our temperature estimates from two different spectral orders agree with the temperatures inferred from the SED (Section \ref{sec: rv}).  Our derived stellar radius from the light curves is well constrained and in good agreement with that derived from the SED and main-sequence evolutionary models as well.  If we were to adopt  (with dubious motivation) the smaller secondary mass of $M_{\rm 2}$=0.22 \msun\ our analysis would yield slightly smaller component radii, but only by $\lesssim$10\%.    

\subsubsection{Out-of-Eclipse Modulations}
\label{sec:outofeclipsevar}
\begin{figure}[h]
    \centering
        \includegraphics[width=1.0\linewidth]{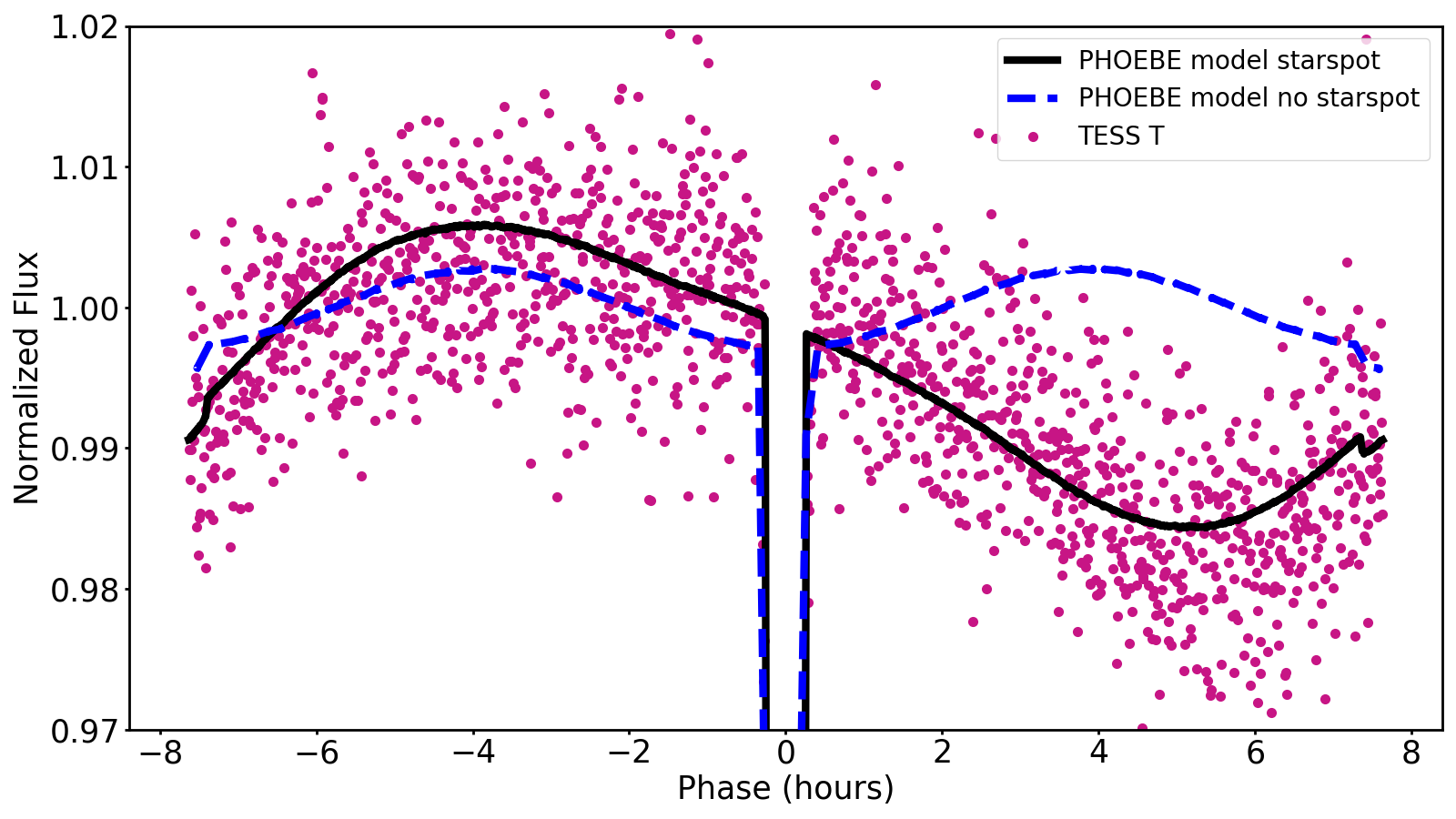}
    \caption{TESS S51 phase-folded light curve (\textit{magenta points}) with best-fitting model including Roche distortion and irradiation effects \textit{(blue dashed curve)}. A  model including the predicted Roche distortion effects and a cool starspot provides a better match to the data \textit{(black solid curve)}. In either case, the model predicts a very shallow secondary eclipse.}
    \label{fig:spottedmodel}
\end{figure}

Figure \ref{fig:spottedmodel} shows the full-period phase-folded TESS light curve \textit{(magenta points)} and the model light curve based on the best-fitting parameters \textit{(dashed line)}. Gravity darkening coefficients were retrieved from atmosphere tables within PHOEBE. The nominal model geometry predicts a secondary eclipse of depth $\sim$0.1\% near +7.63 hours after primary eclipse, but the eclipse depth is small compared to the noise in the data. The ellipsoidal variations predicted by Roche distortion in the nominal best-fitting model are \textless1\% full amplitude and doubly periodic. This model is inconsistent with the light curve, which shows singly periodic modulations. Irradiation and reflected light effects, are insufficient to explain this discrepancy, as discussed in Section \ref{sec: lc}. Instrumental effects would also not phase with the period of the eclipse. 

We attempted to model the out-of-eclipse variations by adding a single cool spot on the M-dwarf to the nominal Roche distorted model. The solid black curve in Figure \ref{fig:spottedmodel} depicts this spotted model, which includes a large equatorial spot with a 55$\degree$ radius at 70$\degree$ longitude (i.e., on the trailing face of the M dwarf) with a temperature 0.995 that of the star. This model approximates the out-of-eclipse light curve, including the singly periodic modulation. We interpret this improved agreement to indicate that additional features such as a quasi-permanent phase-locked (over at least the 23 day TESS Sector) star spot is needed to match the TESS S51 data. In reality, the star may host numerous star spots, but our single-spot model is the simplest possible model for illustrative purposes. The presence of spots is not unexpected, given the high level of stellar activity indicated by the emission lines present in the LAMOST spectrum (Figure~\ref{fig:lamostspectrum}) and the discovery of stellar flares. 

% \begin{table}[h]
%     \centering 
%     \caption{Spot Parameters}
%     \begin{tabular}{c c c}
%     \hline
%     \hline
%     Longitude & Radius & Relative Temperature\cr
%     (deg) & (deg) &  \cr
%     \hline
%     70 &  55 & 0.995 \cr
%     \hline
%     \end{tabular}  
%     \label{tab: spot parameters} 
% \end{table}
% \FloatBarrier

\subsection{Stellar Activity Indicators}
\subsubsection{Emission \& Flare Event}

\begin{figure}[h]
    \centering
    \includegraphics[width=1.0\linewidth]{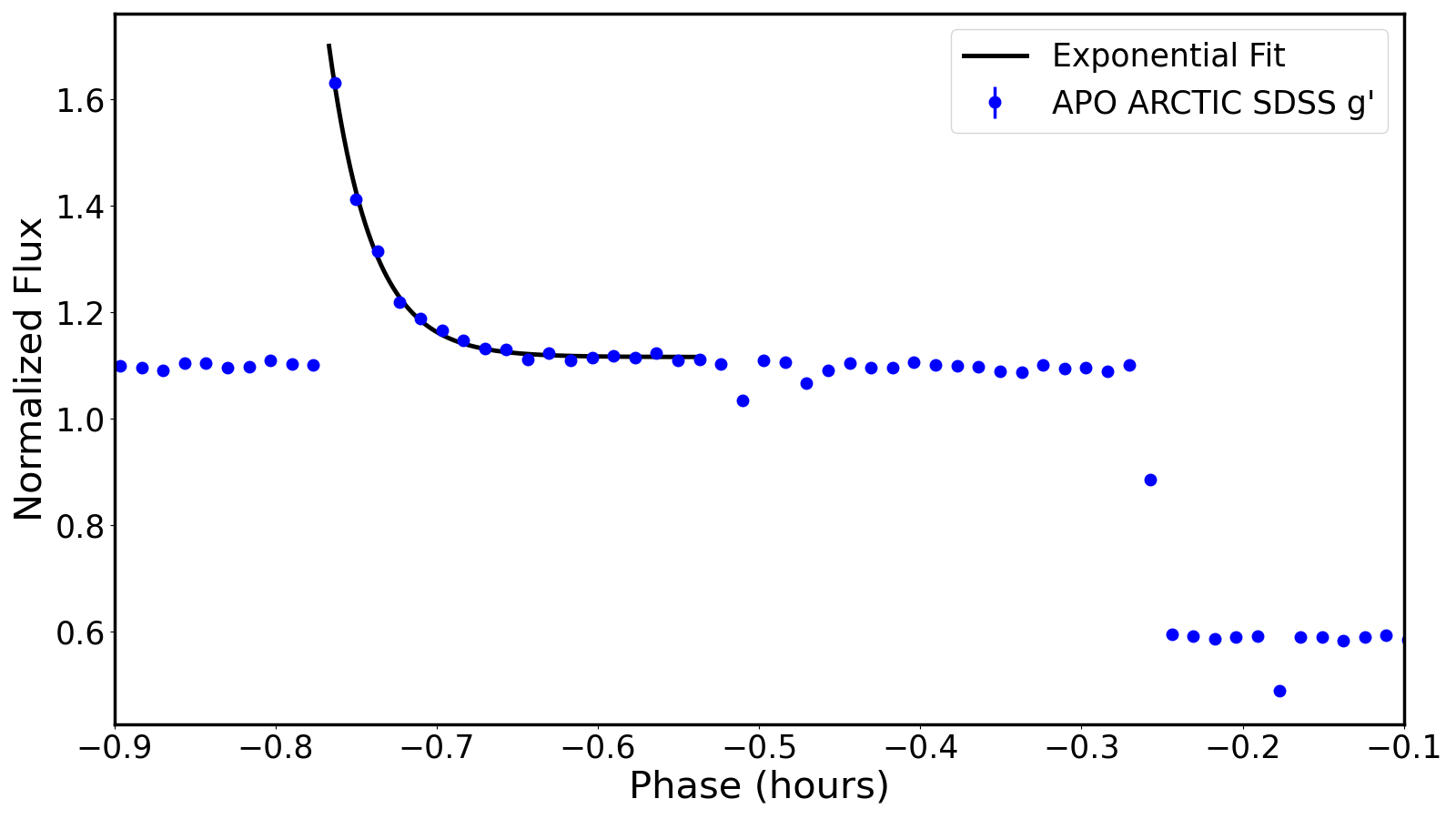}
    \caption{Pre-ingress g$^\prime$ light curve from ARCTIC at APO showing a flare event, fitted as an exponential with a decay timescale of 1.6 minutes \textit{(black curve)}.}
    \label{fig:apoflare}
\end{figure}
% \FloatBarrier

The emission lines of H$\alpha$ and \ion{Ca}{2} H \& K in the LAMOST spectrum (Figure \ref{fig:lamostspectrum}) indicate stellar activity. Another sign of stellar activity is flaring, a common phenomenon in late-type M-dwarf stars \citep{hawley_kepler_2014, gunther_stellar_2019, sethi_tight_2024}. Figure \ref{fig:apoflare} shows the pre-ingress light curve in the 45-s cadence photometric data obtained in the SDSS g$^\prime$ band on ARCTIC plotted in blue. In addition to the transit dip, seven data points show a large excess in flux (up to 60\% above the baseline) that we interpret as a stellar flare. An exponential fit to the flare is plotted in black. The flare's maximum at BJD 2460786.8796 is at least $\sim$63\% higher than the quiescent brightness of the system in the SDSS g$^\prime$ passband and shows a decay timescale of $\sim$1.6 minutes. Although the timing of the flare does not allow for the determination of the physical location, the flare shape is consistent with an M-dwarf flare, having a rapid rise and near-exponential decay \citep{law_three_2012}. 

\subsection{Radial Velocity Curve}
\label{sec: rv} 
We fit a dynamical \texttt{PHOEBE} radial velocity curve to the 12 radial velocity measurements using a similar MCMC method as described in Section \ref{sec: lc}, fixing inclination $i$=89.0\degree$\pm$0.4 from the best-fitting light curve and allowing $M_1$, eccentricity ($e$), and systemic velocity ($\gamma$) in the barycentric velocity frame as free parameters. The mass of the M-dwarf was fixed\footnote{This system is a single-lined spectroscopic binary, where only spectral features from the M-dwarf are detected.} at \msec=0.34$\pm$0.02 based on a fit to the MIST \citep{dotter_mesa_2016, choi_mesa_2016} stellar evolutionary models using \texttt{EXOFASTv2} \citep{eastman_exofastv2_2019} as constrained by the $Gaia$ parallax and eight broadband  optical/infrared flux measurements (Pan Stars g$^\prime$ through WISE2).  The most probable radius derived from the SED is \rsec=0.34$\pm$0.02 \rsun, in excellent agreement with the radius obtained from the light curve above.  The most probable temperature derived from the SED is \teff=3182$\pm$36~K, in good agreement with the temperature determined independently from the HPF spectra.  The mass and radius lead to $\log g$=4.94$\pm$0.02. This agreement with standard evolutionary models supports the proposition that the secondary in this system has main-sequence characteristics.  

Table \ref{tab: final parameters} provide best-fitting \mprim=0.61$\pm$0.04, right around the peak of the WD mass function \citep{mccleery_gaia_2020, bergeron_spectroscopic_1992}. Table \ref{tab: orbital parameters} lists best-fitting and one-$\sigma$ error bars for eccentricity $e = 0.001\pm0.001$, $M_1 = 0.61\pm0.04$, and $\gamma=-12.2\pm0.2$ km s$^{-1}$. As with the light curve fit, the posterior distribution is, in most cases, a tightly constrained Gaussian for each free parameter. The zero eccentricity is expected for a short-period system \citep{goldman_orbital_1991, mayor_orbit_1984}. $\omega$ is not fit owing to the low eccentricity of the system. 

\begin{figure}[h]
    \centering
    \includegraphics[width=1.0\linewidth]{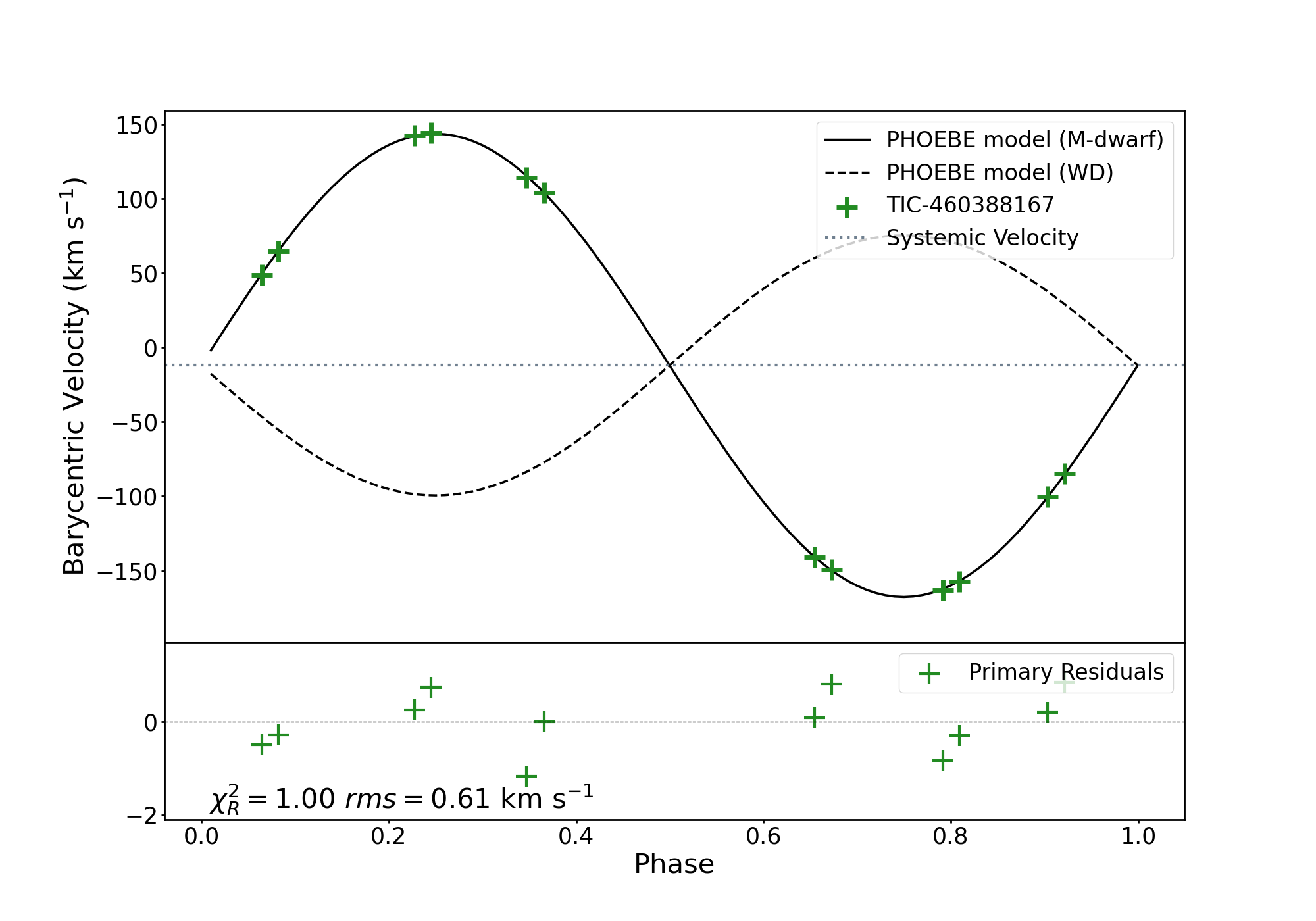}
    \caption{Phase-folded radial velocity measurements \textit{(green +'s)}, and best-fitting \texttt{PHOEBE} model plotted in \textit{(black curve)} \textit{(top panel)}. Residuals \textit{(bottom panel)}.}
    \label{fig:radialvelocities}
\end{figure}

Figure \ref{fig:radialvelocities} shows the radial velocity measurements \textit{(green +'s)} and best-fitting \texttt{PHOEBE} model \textit{(black curve)}. Error bars are estimated based on the Gaussian peak uncertainty to be $\sim$0.2 \kms, but are too small to be seen in the plot. The radial velocity curve is phase-folded corresponding to the best fitting $t_0$ and period found in the light curve analysis. The lower panel of Figure \ref{fig:radialvelocities} shows residuals. Text within the panel states reduced chi-squared and rms. The model reveals that the orbit is nearly circular and has a semi-amplitude of $\sim$150 km s$^{-1}$, which was the first conclusive evidence that the companion to \tic B was massive and stellar in nature.
%The large reduced $\chi ^2$ suggests that the formal uncertainties do not capture the entire uncertainty budget.  Stellar variability/activity may be responsible for the significant RV residuals.

\subsection{Broadening Function}
\begin{figure}[h]
    \centering
    \includegraphics[width=1.0\linewidth]{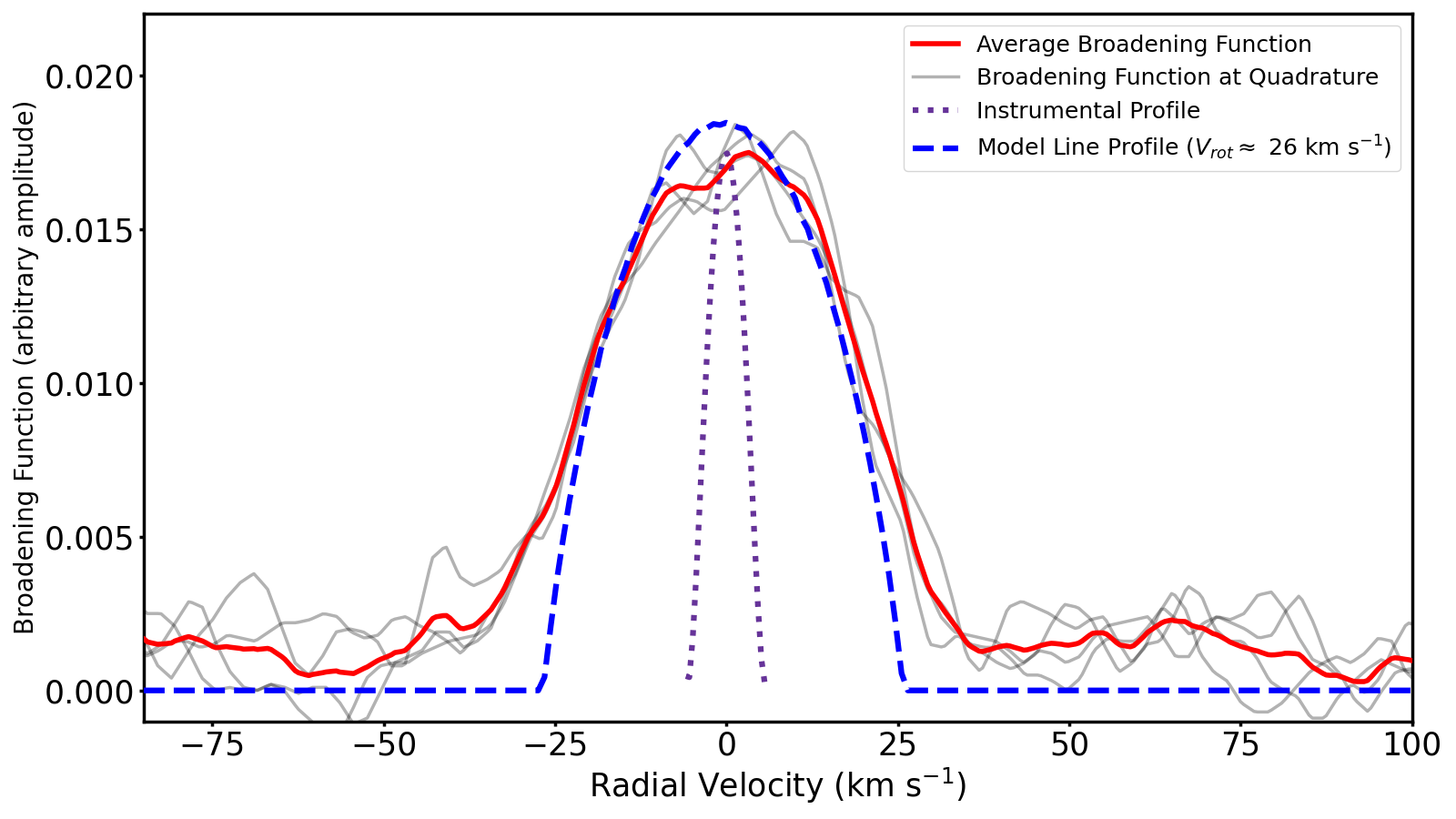}
    \caption{Broadening functions from 4 HPF spectra (phases 0.212, 0.231, 0.776, 0.795) are plotted in black, with the average broadening function plotted in red and model line profile overplotted in blue dashed line. The expected synchronous rotational profile is plotted in the purple dashed line.}
    \label{fig:avgrotprof}
\end{figure}

We computed a broadening function for each of the 12 HPF spectra as described in Section \ref{sec:hpfspec}. Figure \ref{fig:avgrotprof} shows the four broadening functions around phase 0.2 (from BJD 2460734.89995 and 2460734.91192) shifted to a common center \textit{(gray)}, the average broadening function \textit{(red)}, and the \texttt{PHOEBE} best-fitting rotational profile \textit{(blue)}. The instrumental profile of the HPF instrument \textit{(purple dotted line)} is very narrow---$\approx$ 5 \kms\ FWHM---as discerned from an HPF spectrum of the slowly rotating M dwarf Barnard's Star (\textit{green dotted line}), so it does not contribute measurably to the measured rotational profile. Phases near 0.2 were selected due to the minimal effects orbital broadening. The \texttt{PHOEBE} model assumes synchronous rotation due to the short period of \tic\, and included finite time of integration sampling to account for the effects of orbital smearing during the 969 s exposures. The expected maximum rotational velocity assuming synchronous rotation \textit{(dashed blue)} corresponds to $V\sin i \approx V_{rot}$ $\approx$ 26 km s$^{-1}$ based on the well-constrained orbital period and stellar radius. We present the first velocity resolved profile of the main-sequence companion in a PCEB\footnote{\cite{bleach_measurement_2002} measured $V\sin i$ for two PCEBs, but since these systems are not eclipsing the true rotational profile is unknown.}, showing the best-fitting model of rotational line profile was a good match to the measured line profile. This confirms that \tic B exhibits synchronous rotation, which is expected for all systems with orbital periods $P\lesssim$2 d \citep{gladman_synchronous_1996, mathieu_circularized_1988}. 
% Zahn 1992 and 1977

% \FloatBarrier
%add residuals
\subsection{Stellar and Orbital Parameters Summary}
The stellar and orbital parameters for \tic\ are listed in the Tables \ref{tab: final parameters} and \ref{tab: orbital parameters}.

\begin{table}[h]
\centering
\caption{Stellar Parameters}
\begin{threeparttable}
\begin{tabular}{c c c}
\hline\hline
Parameter & WD  & M dwarf  \\
\hline
$T$ (K) & ${7607}\pm128$\tnote{a} & ${3151}\pm59$\tnote{c} \\

$R_{*}$ (R$_{\odot}$) & $0.0131\pm0.0004$\tnote{a} & $0.327\pm0.006$\tnote{a} \\

$M$ (M$_{\odot}$) & $0.61\pm0.04$\tnote{b} & $0.34\pm0.01$\tnote{d} \\

$\log g$ & $7.991\pm0.033$\tnote{a} & $4.94\pm0.02$\tnote{c} \\

$V_{\rm rot}$ (km s$^{-1}$) & \nodata & $26.0\pm0.4$\tnote{b} \\

$[{Fe/H}]$ & \nodata & $-0.04\pm0.16$\tnote{c} \\
\hline
\end{tabular}

\begin{tablenotes}
\footnotesize
\item[a] Derived from light curve.
\item[b] Derived from radial velocity.
\item[c] Derived from HPF Spectra.
\item[d] Derived from SED fitting.
\end{tablenotes}

\end{threeparttable}
\label{tab: final parameters}
\end{table}

\label{sec: orbital parameters}

\begin{table}[h]
\centering
\caption{Orbital Parameters}
\begin{threeparttable}
\begin{tabular}{c c}
\hline\hline
Parameter & TIC-460388167 \\
\hline
$P$ (d) & $0.63596258 \pm 0.00000012$\tnote{a} \\

$t_0$ (BJD) & $2460786.91153 \pm 0.00034$\tnote{a} \\

$t_s$ (BJD) & $2459701.00542 \pm 0.00027$\tnote{a} \\

$i$ (deg) & $89.0 \pm 0.4$\tnote{a} \\

$e$ & $0.001 \pm 0.001$\tnote{b} \\

$\gamma$ (km s$^{-1}$) & $-12.2 \pm 0.2$\tnote{b} \\

$q$ & $0.557 \pm 0.025$\tnote{b} \\

$a$ ($R_{\odot}$) & $3.1$\tnote{b} \\

$K_2$ (km s$^{-1}$) & $156.2 \pm 3.1$\tnote{b} \\

$K_1$ (km s$^{-1}$) & $87.1 \pm 2.3$\tnote{b} \\
\hline
\end{tabular}

\begin{tablenotes}
\footnotesize
\item[a] Derived from light curve.
\item[b] Derived from radial velocity.
\end{tablenotes}

\end{threeparttable}
\label{tab: orbital parameters}
\end{table}

% \FloatBarrier

\subsection{Spectral Energy Distribution Analysis}

\begin{figure}[h]
    \centering
        \includegraphics[width=1.0\linewidth]{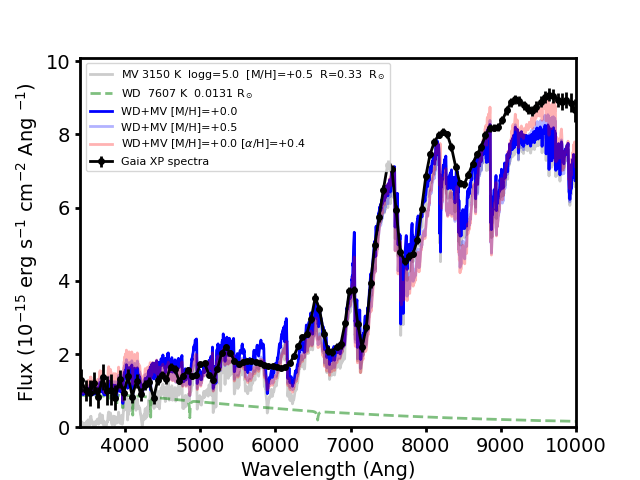}
    \caption{SED of the \tic\ system from $Gaia$ XP spectra (\textit{black}). Colored curves depict stellar atmosphere models at the distance of the system using the best-fitting stellar parameters determined from the light curve analysis, as indicated by the legend.  The model system matches the data well with no free parameters.  }
    \label{fig:SED}
\end{figure}

Figure~\ref{fig:SED} shows the spectral energy distribution (SED) of the \tic\ system over the wavelength range 3400--10,000 \AA\ using the $Gaia$ XP spectra \citep[\textit{black filled circles and error bars},][]{Gaia_2023A&A...674A...1G}. A green dashed curve depicts a model white dwarf atmosphere \citep{koester_white_2010} with the size and temperature determined from the light curve analysis at a distance of 54 pc. The light gray curve shows a ``New Era'' Phoenix \citep{hauschildt_newera_2025} model atmosphere for a $T_{\rm eff}$=3150~K (interpolated between models and smoothed to the resolution of the Gaia XP spectra), $\log g$=5.0, [M/H]=0.0 star of the radius determined in the light curve analysis.  Blue/faded blue/faded red curves depict the composite model WD+MV system for three different metallicities [M/H]=0.0/[M/H]=+0.5/[$\alpha$/H]=+0.4, as indicated in the legend. The differences between these three models are minimal.  Figure~\ref{fig:SED} demonstrates that all three model SEDs provide a good match to the data, reproducing the overall flux across the optical range and matching most of the broad spectral features with no free parameters.  There are notable discrepancies in the range 5500--6100~\AA, where the data are remarkably flat compared to the models.  None of the available models can reproduce this flatness in this regime.   At $>$8800~\AA\ the models all show a deficit relative to the data.  Adopting adjacent models at slightly different $\log g$ or metallicity does not reduce these discrepancies.  However, the discrepancies are similar in five other randomly selected $\approx$3150~K dwarf stars having $Gaia$ XP spectra, so we conclude that these are real characteristics common to cool stars that depart systematically from the Phoenix models rather than anomalous features unique to \tic B.         

% \FloatBarrier

\section{Discussion}
\label{sec:discussion}

\subsection{WDMS Binary Demographics}

\begin{figure}[ht]
    \centering
        \includegraphics[width=1.0\linewidth]{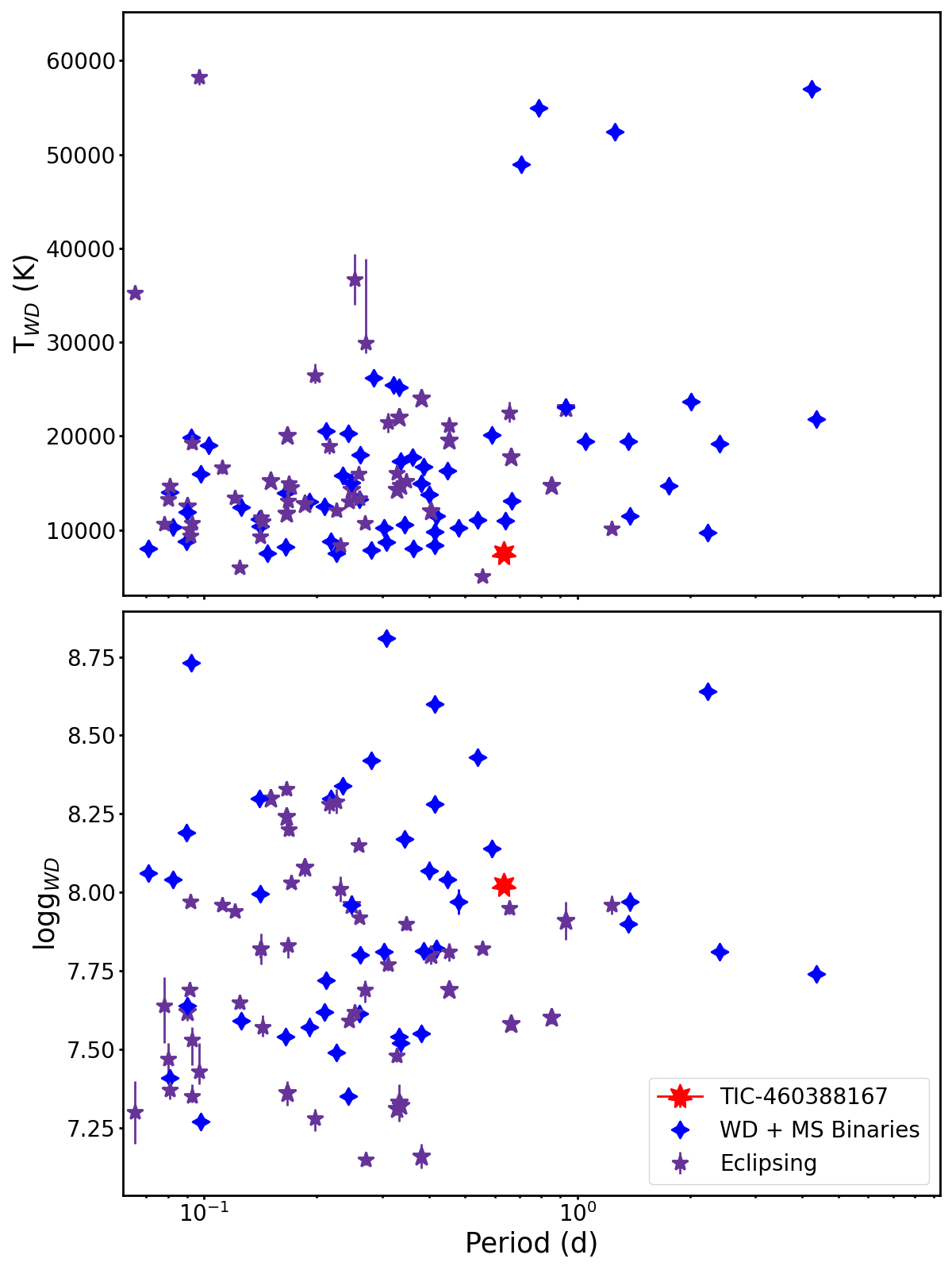}%{figure10.png}
    \caption{WD temperature \textit{(upper panel)} and WD \logg\ \textit{(lower panel)} vs. orbital period. The red seven-pointed star depicts \tic A, the blue four-pointed stars shows other WD+M-dwarf binaries, and the purple five-pointed stars show the WDMS binaries which are eclipsing.}
    \label{fig:popcomp2}
\end{figure}
% \FloatBarrier

Figure \ref{fig:popcomp2} plots WD temperature vs. orbital period \textit{(upper panel)} and WD \logg\ vs.~orbital period \textit{(lower panel)} for \tic A \textit{(red seven-pointed star)} and WDMS binaries from the literature \citep[\textit{blue four-pointed stars},][]{rebassa-mansergas_magnitude-limited_2025,inight_towards_2021}, with the eclipsing WDMS systems \citep[\textit{purple five-pointed stars},][]{pyrzas_post-common_2012,parsons_deeply_2011,rebassa-mansergas_magnitude-limited_2025,brown_characterising_2022}. The vast majority of WDMS binaries lie in the period range of 0.1--10 days. Most known systems have temperatures higher than $\sim$8000 K. This is a selection effect, since cool WDs are faint and more difficult to detect. At $T_{\rm eff}$=7600~K, \tic\ hosts one of the coolest WDs (in the 97th percentile)\footnote{\cite{ribeiro_accretion_2013} report a WDMS binary RR Caeli with a primary of $T_{\rm eff}\approx7200$ K}. If it were not eclipsing, it would have been difficult to detect on the basis of the SED or spectra alone. With $P$=0.63~d, it is also among the longest period eclipsing systems known (in the 92nd percentile)---systems where orbital and component parameters can be measured with great precision owing to the measured inclination.

\subsection{System Evolutionary Status}
\label{sec:evolstat}

\subsubsection{Mass-Radius-Temperature Relation}
\label{sec:MRR}
We measured the mass and radius of the WD to be $M_1 = 0.61\pm0.04~M_\odot$ and $R_1 = 0.0131\pm0.0004~R_\odot$, and determined the effective temperature to be $T_1 = 7607\pm128$~K. These values are consistent with the mass-radius-temperature relation for WDs \citep[see Fig.~7 in][]{pereiras_mass-radius_2025} and the MRR in \cite{bedard_measurements_2017}, especially given the uncertainty on the WD mass. Our measurements for \tic A can augment the sample of WDs that can be used to constrain theoretical cooling models in the low-temperature limit.

\subsubsection{Age}
\label{sec:age}
We estimate the initial mass of the WD progenitor to have been $\sim$1.4--2 $M_\odot$ based on the initial-final mass relations (IFMR) of \cite{cummings_white_2018} and \cite{el-badry_empirical_2018}, respectively. This indicates the progenitor was a mid-F through mid-A star with an expected MS lifetime of 1.3--3 Gyr based on MIST evolutionary tracks \citep{dotter_mesa_2016, choi_mesa_2016}. Therefore, the system must be at least this old. Based on the temperature and mass of the WD, we use the WD cooling tracks of \cite{bedard_spectral_2020} to determine the WD age to be around 1.2 Gyr. However, this age should only be taken as an approximation, since it is possible the CE phase truncated the WD evolution before it naturally evolved to the tip of the giant branch. Based on the proper motions of the system, we find that it has kinematics consistent with the stars in the Galactic disk rather than the halo, pointing to an origin in the previous $\lesssim$4 Gyr. By comparison, the main-sequence lifetime of the 0.34 \msun\ M dwarf is expected to be 200 Gyr \citep{choi_mesa_2016}, so its nuclear evolution will proceed on a timescale that is long compared to other relevant timescales in the system.  

\subsubsection{Orbital Circularization}
\label{sec:circularization}
We determined the eccentricity of the system to be 0.001$\pm$0.001, which is expected given the 1 Gyr age of the WD. Based on \cite{goldman_orbital_1991,mayor_orbit_1984}, all orbits with periods less than 2 d are circular. The predicted circularization timescale for a binary of this period and mass ratio is $\sim$10$^5$ years \citep[][Eq.~4.43]{hilditch_introduction_2001}, far less than the time since the common envelope phase. 
% It is likely that stellar activity of \tic\ produces a radial velocity jitter that limits the precision of the spectroscopic measurements, a known phenomenon in active stars such as M-dwarfs \citep{lafarga_carmenes_2023, reiners_activity-induced_2009}. This jitter likely causes the appearance of an eccentricity that is larger than the true dynamical value. 
% Hence, the eccentricity of 0.074 should be considered an upper limit.

\subsubsection{Orbital Synchronization}
\label{sec:synchronization}
The rotational synchronization timescale for a binary with a period of 0.636 d is $\sim$3000 years \citep[][Eq.~4.42]{hilditch_introduction_2001}. The rotational broadening profile we measure from the quadrature phase spectra matches the \texttt{PHOEBE} model line profile, computed assuming synchronous rotation, quite well. This is expected given the short period of the system \citep{gladman_synchronous_1996, fleming_rotation_2019}. This is the first directly measured rotational profile of the main-sequence companion in a PCEB system, confirming that the MS component is indeed tidally locked.

\section{Conclusion}
\label{sec: Conclusion}
We have reported the discovery of a new eclipsing PCEB system \tic, utilizing one TESS sector combined with ground-based photometry and spectroscopy. In addition to eclipses, the TESS light curve displayed continuous variability which we modeled as a combination of ellipsoidal variations and star spots. The high--quality light curves have allowed for precise measurements of the radii of both components. The secondary in \tic\ is an active star based on the flaring and emission lines seen in its spectrum, consistent with other mid-M-dwarfs. \tic A is one of the coolest known WDs paired with a MS star. The WD mass of 0.61~$M_\odot$ lies near the peak of the WD mass distribution function. We found that \tic\ does have a potentially observable---although shallow---secondary eclipse (0.1\% in TESS T band) that occurs 7.63 hours after primary eclipse. To our knowledge, we have measured the first radial velocity profile of the main-sequence secondary in a PCEB system like this, representing a direct confirmation that it is tidally locked and rotates at the orbital period, as expected. These component parameters add to the ongoing understanding of PCEB systems and provide new data to understand the impacts of the common envelope phase on its constituent stars.

\section{Acknowledgments}

This study is based on observations obtained with the Hobby-Eberly Telescope (HET), which is a
joint project of the University of Texas at Austin, the Pennsylvania State
University, Ludwig-Maximillians-Universitaet Muenchen, and Georg-August
Universitaet Goettingen. The HET is named in honor of its principal benefactors,
William P. Hobby and Robert E. Eberly.

We acknowledge support from U.S. Contributions to Ariel Preparatory Science NASA grant 80NSSC25K0184.

These results are based on observations obtained with the Habitable-zone Planet
Finder Spectrograph on the HET. The HPF team acknowledges support from NSF
grants AST-1006676, AST-1126413, AST-1310885, AST-1517592, AST-1310875, ATI
2009889, ATI-2009982, AST-2108512, AST-2108801, and the NASA Astrobiology Institute
(NNA09DA76A) in the pursuit of precision radial velocities in the NIR. The HPF
team also acknowledges support from the Heising-Simons Foundation via grant
2017-0494.

This research has made use of the NASA Exoplanet Archive, which is operated by
the California Institute of Technology, under contract with the National
Aeronautics and Space Administration under the Exoplanet Exploration Program.

This work has made use of data from the European Space Agency (ESA) mission
{\it Gaia} (\url{https://www.cosmos.esa.int/gaia}), processed by the {\it Gaia}
Data Processing and Analysis Consortium (DPAC,
\url{https://www.cosmos.esa.int/web/gaia/dpac/consortium}). Funding for the DPAC
has been provided by national institutions, in particular the institutions
participating in the {\it Gaia} Multilateral Agreement.

CIC acknowledges support by NASA Headquarters through (i) an appointment to the NASA Postdoctoral Program at the Goddard Space Flight Center, administered by ORAU through a contract with NASA and (ii) under award number 80GSFC24M0006.

% CIC acknowledges support by NASA Headquarters through an appointment to the NASA
% Postdoctoral Program at the Goddard Space Flight Center, administered by ORAU
% through a contract with NASA.

% Supported by the National Science Foundation under Grants No. AST-1440341 and
% AST-2034437 and a collaboration including current partners Caltech, IPAC, the
% Oskar Klein Center at Stockholm University, the University of Maryland,
% University of California, Berkeley , the University of Wisconsin at Milwaukee,
% University of Warwick, Ruhr University, Cornell University, Northwestern
% University and Drexel University. Operations are conducted by COO, IPAC, and UW.

% We thank Jason Eastman and Noah Vowell for helpful conversations regarding the
% use of EXOFASTv2.

% We thank Maxwell Moe for helpful conversations regarding formation mechanisms
% and biases.

% We thank Brock A. Parker for helpful conversations regarding data reduction and
% the usage of several software. \\

\facilities{HET, HPF, RBO, TESS, Gaia, ExoFOP, WISE, FWLO:2MASS, CTIO:2MASS, PS1, Sloan, APO, TMMT}
%Facilities: HET, RBO, WIYN, TESS, Gaia, ExoFOP,
% FWLO:2MASS, CTIO:2MASS, AAVSO, Sloan, PS1, WISE,
% NESSI, PO:1.2m, ASAS-SN.
\software{AstroImageJ \citep{collins_astroimagej_2017}, EXOFASTv2 \citep{eastman_exofastv2_2019}, barycorrpy \citep{kanodia_python_2018}, HPF-SpecMatch \citep{stefansson_sub-neptune-sized_2020-1}, TESS-Gaia lightcurve \citep{han_tessgaia_2023}, PHOEBE2v2.4 \citep{prsa_physics_2016}, Matplotlib \citep{hunter_matplotlib_2007}, pandas \citep{mckinney_data_2010}, numpy \citep{harris_array_2020}} \\

\newpage

\bibliography{Paper1}{}

\bibliographystyle{aasjournal}

\end{document}